\documentclass[aps,prl,twocolumn,superscriptaddress,floatfix]{revtex4}
\usepackage{graphicx}
\usepackage{amssymb}

\def\BKFA{Ba$_{1-x}$K$_{x}$Fe$_2$As$_2$} 
\def\BFA{BaFe$_2$As$_2$} 
\def\BFCA{Ba(Fe$_{1-x}$Co$_x$)$_2$As$_2$}
\def\LFAOF{LaFeAsO$_{1-x}$F$_x$}

\def\Tc{T$_\mathrm{c}$}
\def\TN{T$_\mathrm{N}$}
\def\Ts{T$_\mathrm{s}$}

\begin{document}

\title{Magnetoelastic Coupling in the Phase Diagram of \BKFA}

\author{S. Avci}
\affiliation{Materials Science Division, Argonne National Laboratory, Argonne, IL 60439, USA}
\author{O. Chmaissem}
\affiliation{Materials Science Division, Argonne National Laboratory, Argonne, IL 60439, USA}
\affiliation{Department of Physics, Northern Illinois University, DeKalb, IL 60115, USA}
\author{E. A. Goremychkin}
\author{S. Rosenkranz}
\author{J.-P. Castellan}
\author{D. Y. Chung}
\author{I. S. Todorov}
\author{J. A. Schlueter}
\author{H. Claus}
\affiliation{Materials Science Division, Argonne National Laboratory, Argonne, IL 60439, USA}
\author{M. G. Kanatzidis}
\affiliation{Materials Science Division, Argonne National Laboratory, Argonne, IL 60439, USA}
\affiliation{Department of Chemistry, Northwestern University, Evanston, IL 60208-3113, USA}
\author{A. Daoud-Aladine}
\author{D. Khalyavin}
\affiliation{ISIS Pulsed Neutron and Muon Facility, Rutherford Appleton Laboratory, Chilton, Didcot OX11 0QX, United Kingdom}
\author{R. Osborn}
\affiliation{Materials Science Division, Argonne National Laboratory, Argonne, IL 60439, USA}
\email{ROsborn@anl.gov}

\begin{abstract}
We report a high resolution neutron diffraction investigation of the coupling of structural and magnetic transitions in \BKFA. The tetragonal-orthorhombic and antiferromagnetic transitions are suppressed with potassium-doping, falling to zero at $x\lesssim0.3$. However, unlike \BFCA, the two transitions are first-order and coincident over the entire phase diagram, with a biquadratic coupling of the two order parameters. The phase diagram is refined showing that the onset of superconductivity is at $x=0.133$ with all three phases coexisting until $x\gtrsim0.24$.
\end{abstract}

\date{\today}

\maketitle

Phase competition is an essential ingredient of superconductivity in the iron arsenides and related compounds. The superconducting phase emerges when antiferromagnetism has been suppressed either by hole or electron doping\cite{Wen:2008p7965,Rotter:2008p11893}, applied pressure\cite{Torikachvili:2008p11685}, or disorder\cite{Wadati:2010p34986}, but the nature of the phase boundary from antiferromagnetism to superconductivity is not universal. In the so-called `1111' system, \LFAOF, it has been reported that there is a sharp first-order transition at $x\sim0.045$ from the antiferromagnetic phase to the superconducting phase, but there are conflicting reports of phase coexistence in isostructural compounds containing other rare earth ions\cite{Zhao:2008p13557,Drew:2009p19227,Sanna:2009p29496}. On the other hand, in the `122' systems with the parent compound \BFA, both hole and electron doping produce a gradual suppression of the antiferromagnetism leading to the onset of superconductivity with some overlap of the two phases. 

Antiferromagnetism is also associated with a structural phase transition from tetragonal to orthorhombic symmetry that occurs at a temperature either just above or coincident with the onset of magnetic order\cite{delaCruz:2008p8095,Pratt:2009p29514,Rotter:2008p11893,Fernandes:2010p32347}. It has been proposed that the structural distortion involves a change in the orbital configuration\cite{Shimojima:2010p32066} producing an electronic nematic phase that is either a precursor of, or is driven by, antiferromagnetic correlations. This has led to considerable interest in the role of possible nematic fluctuations in the normal phase of the nominally tetragonal superconductors\cite{Fang:2008p8200,Chuang:2010p31929,Fernandes:2010p34989}. Investigations of the interplay of magnetism, orbital order, and superconductivity are therefore important in unravelling the origin of unconventional superconductivity in these compounds.

When investigating the phase diagram of doped materials, it is a challenge to separate effects due to chemical inhomogeneity from those due to intrinsic phase separation\cite{Mukhopadhyay:2009p31071}. In the `122' compounds, comparisons of bulk diffraction with local probes, such as NMR and $\mu$SR, have led to two different conclusions for the electron-doped compounds, \BFCA, and the hole-doped compounds, \BKFA. In the case of electron-doping, there is evidence of true phase coexistence in the underdoped compounds, with a coupling of the antiferromagnetic and superconducting order parameters\cite{Pratt:2009p29514,Julien:2009p30867}. On the other hand, in the case of hole-doping, local probes have indicated that there may be phase separation  \textit{i.e.}, the antiferromagnetic and superconducting phases occur in separate mesoscopic domains within the crossover region\cite{Julien:2009p30867,Park:2009p20194}. A theoretical analysis of this phase competition concludes that both phase diagrams can be consistent with a superconducting order parameter of $s_{\pm}$ symmetry\cite{Fernandes:2010p32347,Mazin:2008p11687}, whether there is true phase coexistence below a tetracritical point or phase separation close to a first-order bicritical line.

In this paper, we report a reexamination of the phase diagram of \BKFA, one of the most challenging of the iron pnictide superconductors to synthesize. Discrepancies in the published phase diagrams, with antiferromagnetism being suppressed at dopant concentrations varying from $x=0.25$\cite{Johrendt:2009p28737} to 0.4\cite{Chen:2009p15806}, reflect the difficulty of controlling the stoichiometry owing to the high volatility of potassium. Because of this, most research has been conducted on \BFCA\ and other transition-metal-doped compounds. Nevertheless, it is important to study \BKFA, partly to investigate any assymmetry between electron and hole doping in the phase diagram, but also because potassium substitution is intrinsically cleaner, since there is no disorder in the superconducting Fe$_2$As$_2$ planes themselves. By optimizing the homogeneity of potassium-doped samples, we have been able to show that the superconducting phase starts at $x=0.133\pm0.002$ with evidence of phase coexistence, rather than phase separation, up to $x\sim0.24$. Using high-resolution neutron powder diffraction, we observe that the structural and antiferromagnetic transition temperatures are coincident and first-order over the range $0\leqslant x \leqslant 0.24$, with biquadratic coupling at all $x$, a highly unusual form of magnetoelastic coupling that has implications for the nature of the ordered state.

\begin{figure}[!tb]
\centering
\vspace{-0.25in}
\includegraphics[width=\columnwidth]{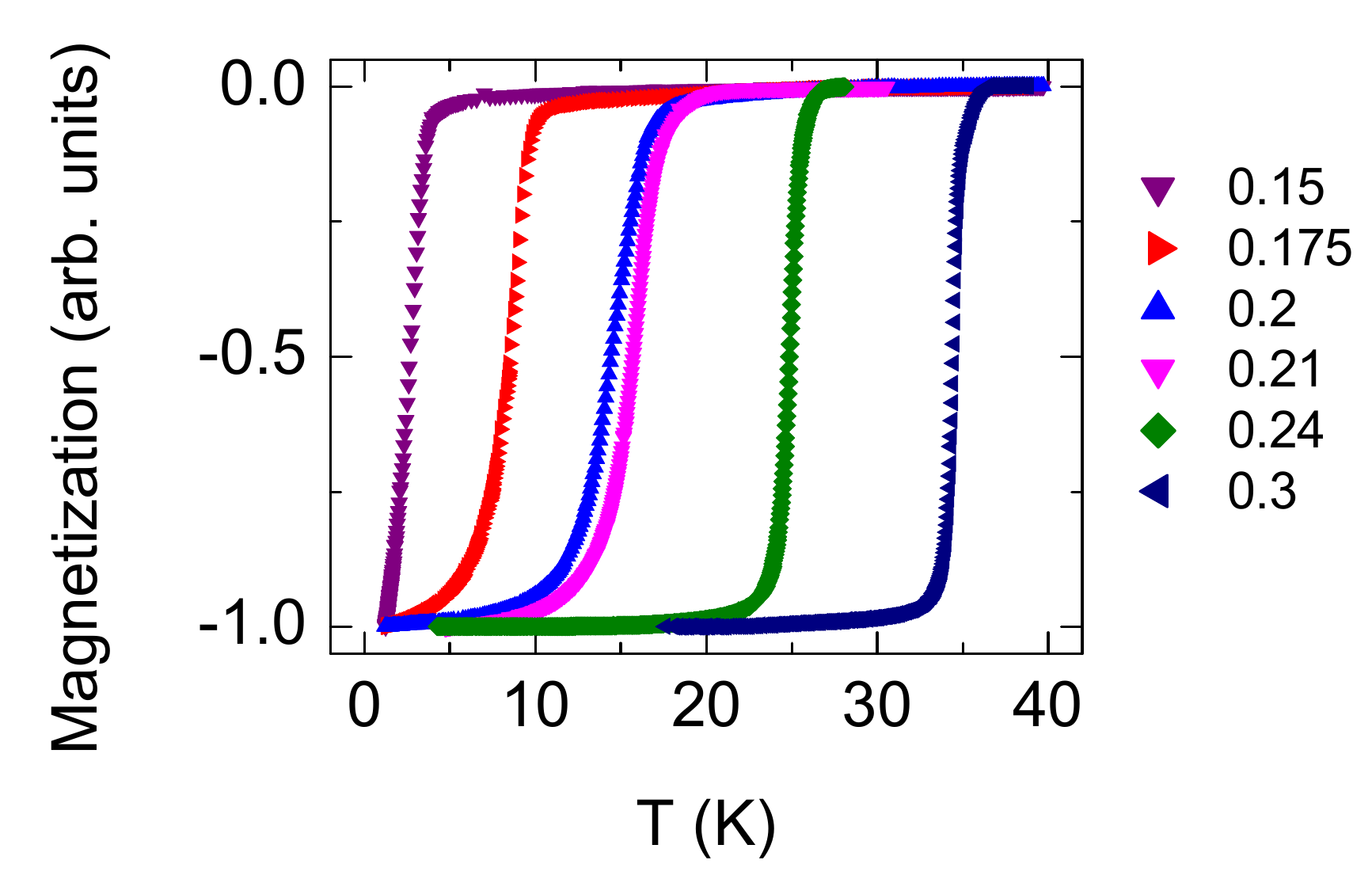}
\caption{Magnetization of \BKFA\ for $x=0.15$, 0.175, 0.2, 0.21, 0.24, and 0.3, measured using a SQUID magnetometer. Samples with $x<0.15$ showed no superconductivity above a temperature of 0.3\,K.
\label{Fig1} }
\vspace{-0.25in}
\end{figure}

In order to overcome the high vapor pressure and reactivity of potassium metal and the formation of more stable K/As binary by-products in the synthesis of \BKFA, we examined all reasonable combinations of reaction parameters (\textit{e.g.}, starting materials, reaction containers, temperature, and heating times, \textit{etc.}) before establishing the optimal conditions to produce high quality homogeneous samples with sharp magnetic and superconducting transitions. Samples were prepared using a stoichiometric mixture of binary BaAs, KAs, and FeAs powders prepared in a N$_2$-filled glove box. The mixtures were loaded in alumina tubes and pre-heated at 500 - 800$^\circ$C.  The pre-annealed mixtures were then ground and loaded in niobium, which were then placed inside quartz tubes. Heating the materials at 1000$^\circ$C for 24 to 48\,h followed by cooling to room temperature over 12 hours resulted in black polycrystalline powders.   Homogeneity of the samples was ensured by repeating this process multiple times.  X-ray diffraction, magnetic susceptibility, and ICP elemental analysis were all used to control and monitor the progress of the sample quality during and after synthesis. High quality samples were successfully synthesized to cover the entire phase diagram of the \BKFA\ series from $0\leqslant x\leqslant1.0$, with increments of $\Delta x=0.025$ from $0.1\leqslant x\leqslant0.25$, close to the superconducting phase boundary. 

The neutron powder diffraction measurements were carried out on the High Resolution Powder Diffractometer (HRPD) at the ISIS Pulsed Neutron Source, whose resolution of 10$^{-4}$ is extremely sensitive to inhomogeneous line-broadening. The high quality of our samples was demonstrated by the constant width of reflections in both undoped and doped compounds, \textit{e.g.} FWHM$\sim0.0037(3)$\,\AA\ for the (220) peak. SQUID (Quantum Design) magnetization measurements were used to determine the superconducting transition temperatures, \Tc, and the N\'eel temperatures, \TN. The peak in the first derivative of the magnetization produced values of \TN\ that were in good agreement with the neutron diffraction measurements over the entire phase diagram. 

\begin{figure}[!b]
\centering
\includegraphics[width=0.8\columnwidth]{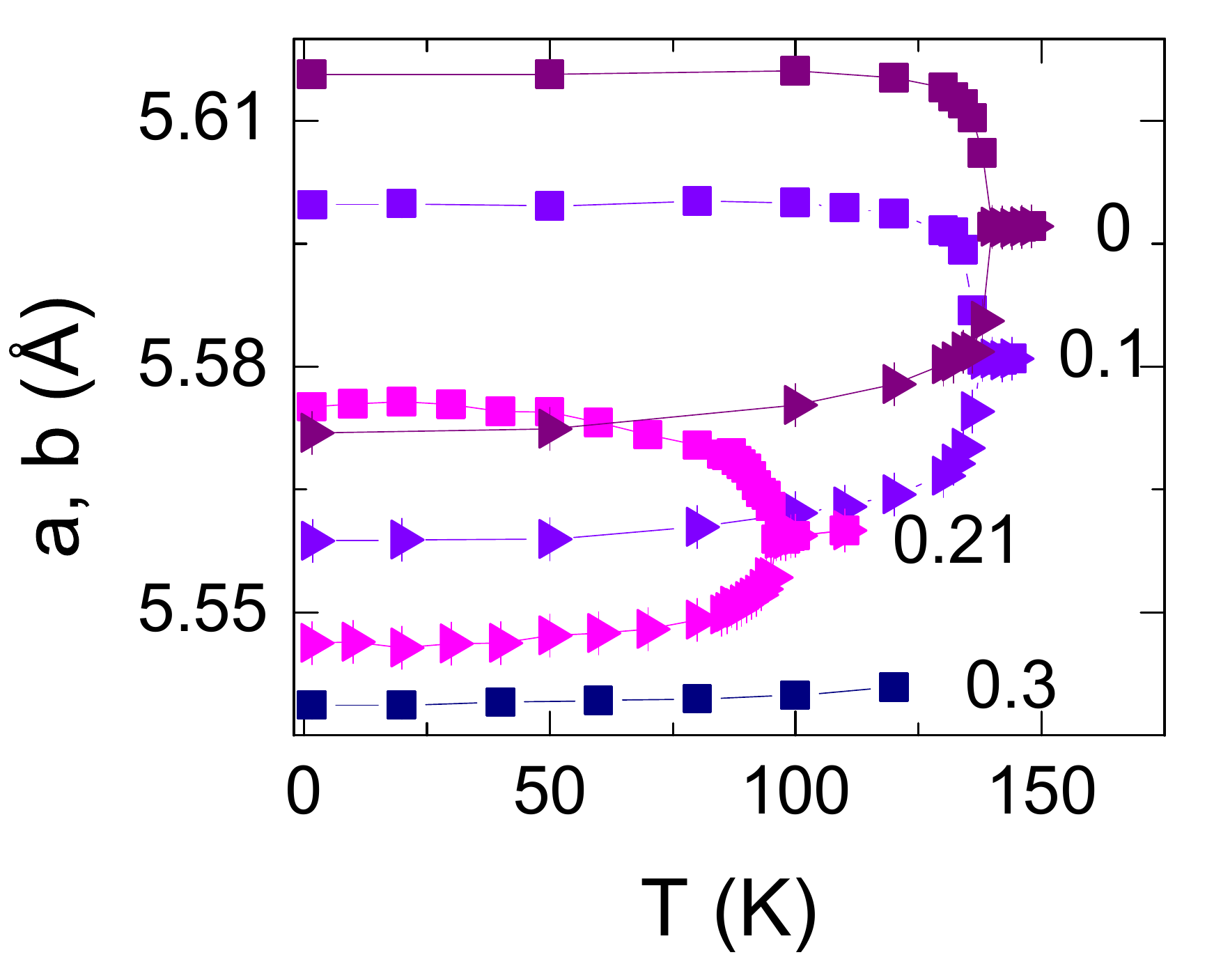}
\vspace{-0.25in}
\caption{Variation of lattice constants $a$ and $b$ with temperature in \BKFA for $x=0$, 0.1, 0.21 and 0.3.
\label{Fig2} }
\vspace{-0.25in}
\end{figure}

The magnetization measurements showed no evidence of superconductivity above 300\,mK for any of the samples with $0 \leqslant x \leqslant 0.125$.  Bulk superconductivity is first seen at $x=0.15$ with a \Tc\ of 4\,K, and then increases more rapidly with potassium concentration than previously seen, peaking at 38\,K for $x=0.4$ before it decreases again to 3\,K for the end member, KFe$_2$As$_2$. The magnetization of the underdoped compounds in Fig. 1 shows that well-defined superconducting transitions are observed even when \Tc\ is varying rapidy with $x$, where the results would be most sensitive to composition fluctuations. Using linear regression of the underdoped region, we estimate the critical concentration for superconductivity to be $x=0.133\pm0.002$. The complete phase diagram is discussed later.

\begin{figure}[!t]
\centering
\includegraphics[width=\columnwidth]{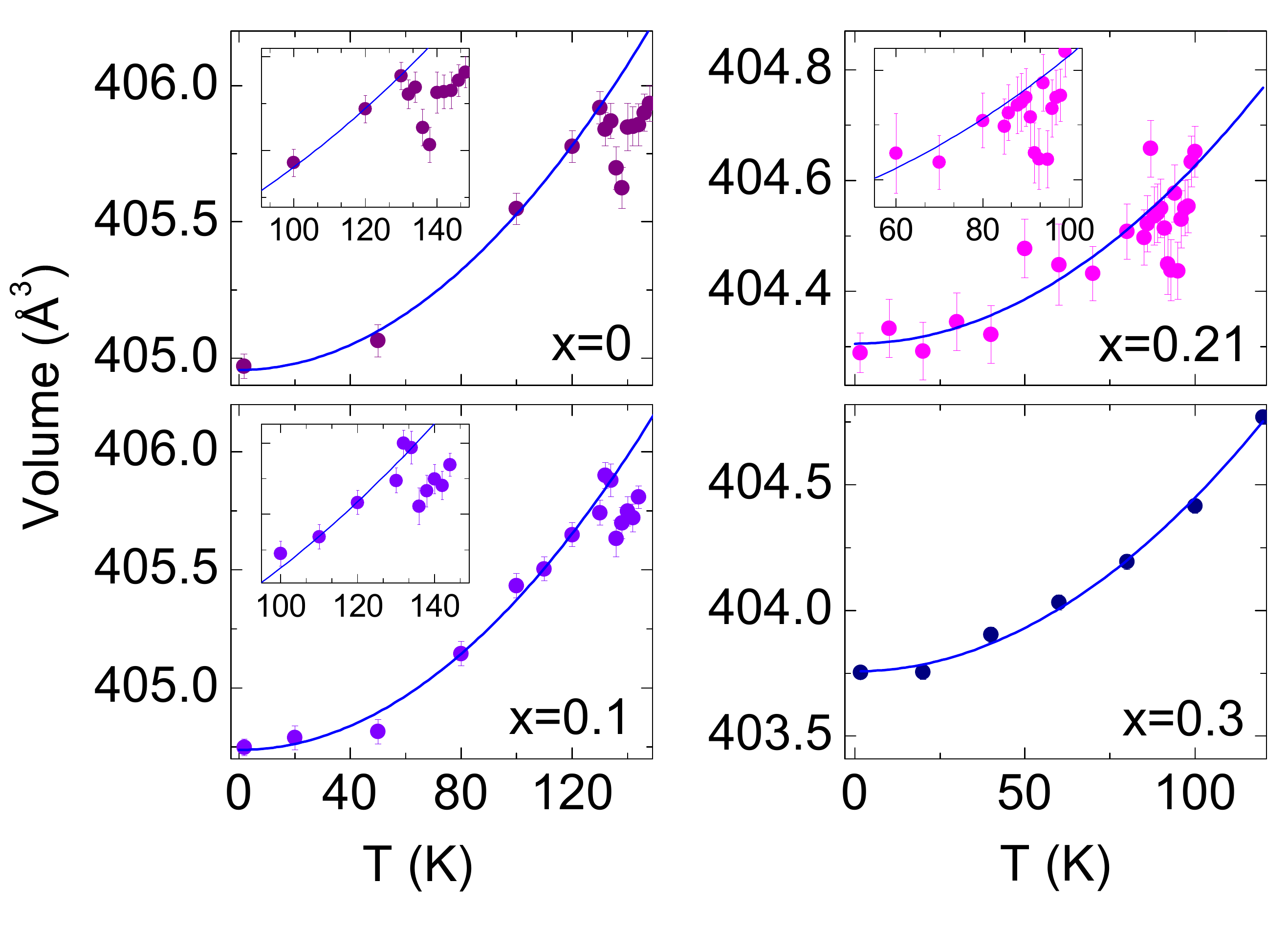}
\vspace{-0.25in}
\caption{Temperature dependence of unit cell volumes for \BKFA\ with $x=0$, 0.1, 0.21 and 0.3. The solid lines are fits below \Ts\ to the quadratic temperature dependence typical of conventional thermal expansion, which is obeyed for $x=0.3$. The insets magnify the region close to \Ts.
\label{Fig3} }
\vspace{-0.25in}
\end{figure}

Rietveld refinements of \BKFA\ confirmed the earlier reports of a structural transition from the tetragonal ThCr$_2$Si$_2$-type structure of space group $I4/mmm$ to the orthorhombic symmetry of the $\beta$-SrRh$_2$As$_2$-type structure of space group $Fmmm$\cite{Rotter:2008p12981}. The structural transition temperature, \Ts\, decreases with potassium doping from 140\,K at $x=0$ to 80\,K at $x=0.24$, and is completely suppressed at $x=0.3$, below the value reported by Chen \textit{et al}\cite{Chen:2009p15806}, but in reasonable agreement with Johrendt \textit{et al}\cite{Johrendt:2009p28737}. Fig. 2 shows the temperature dependence of the orthorhombic splitting for $x=0$, 0.1, and 0.21, and the absence of any splitting for $x=0.3$. 

The high $d$-spacing resolution on HRPD allows extremely small volume anomalies to be observed at \Ts\ for all values of $x$ (Fig. 3), a clear signature that the structural phase transitions are first-order.  Although the equivalent transitions were also observed to be first-order in SrFe$_2$As$_2$\cite{Jesche:2008p14067} and CaFe$_2$As$_2$\cite{Goldman:2008p13171}, a previous neutron study concluded that the transition  in \BFA\ was second-order with 3D critical fluctuations above \Ts\ and an anomalously small 2D critical exponent of $\beta=0.103$ below\cite{Wilson:2009p23312}. They attributed this behavior to a 3D to 2D crossover in the immediate vicinity of the transition. However, they did not rule out that the transition was weakly first-order and subsequent x-ray and heat capacity measurements on a sample prepared with longer annealing times identified a small first-order jump\cite{Rotundu:2010p35102}. The HRPD data provide clear evidence that the transition is first-order and that this characterizes the transition over the entire phase diagram. 

The neutron powder diffraction data also reveals the presence of weak magnetic Bragg reflections below the N\'eel temperatures for all the orthorhombic samples. The peaks indexed as (121) and (103), with $d$-spacings of 2.45\,\AA\ and 3.43\,\AA\ respectively, are consistent with the previously identified spin density wave order\cite{Rotter:2008p12981}. The magnetic structure was refined using the symmetry of the magnetic space group $F_{c}mm^\prime m^\prime$.  In this model, the removal of time reversal symmetry from the last two mirror planes (perpendicular to the $b$ and $c$ axes) resulted in an arrangement in which the Fe magnetic moments are antiferromagnetically coupled along the $x$ and $z$ direction but ferromagnetically coupled along the $y$ axis. The Fe magnetic moment refines to 0.75(3)\,$\mu_B$ at 1.7\,K for the parent \BFA\ material.  A linearly decreasing magnetic moment was observed upon increasing the K content from $x=0.1$ to $0.24$.  No magnetic peaks are observed beyond this limiting value.  

\begin{figure}[!b]
\centering
\includegraphics[width=\columnwidth]{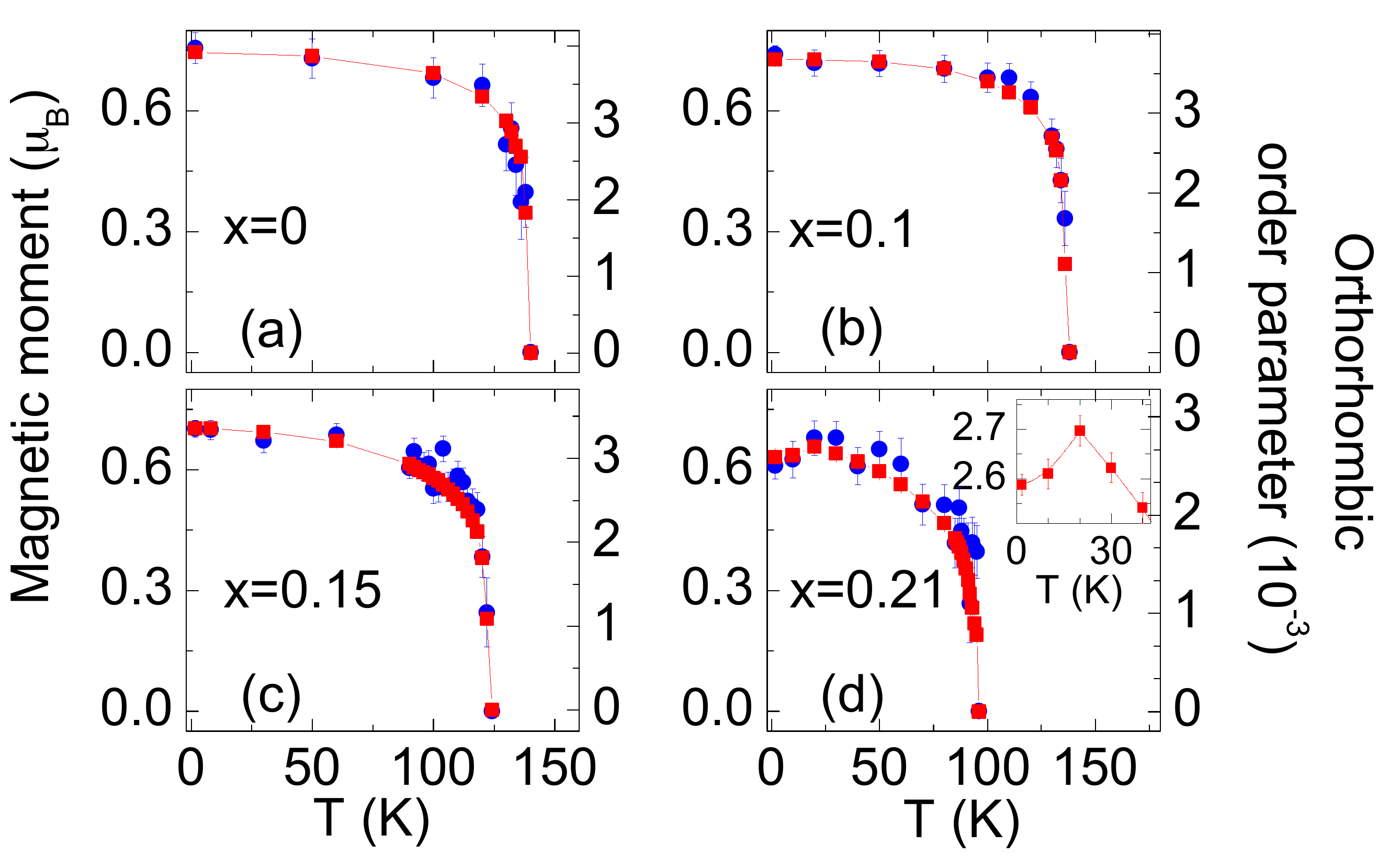}
\vspace{-0.2in}
\caption{Refined magnetic moments (blue circles) and orthorhombic order parameter (red squares) as a function of temperature for x=0, 0.1, 0.15 and 0.21 samples. Solid lines are guide to the eye.
\label{Fig4} }
\vspace{-0.2in}
\end{figure}

Fig. 4 shows a comparison of the temperature dependence of the refined magnetic moment and the orthorhombic order parameter, defined by the expression $\delta=(a-b)/(a+b)$, where $a$ and $b$ are the in-plane orthorhombic lattice parameters. Although the statistical precision of the magnetic order parameter, $M$, is much less than the orthorhombic order parameter, it is clear that they have identical temperature dependences at all compositions. The data are in clear contradiction to an earlier NMR report that the two transitions are distinct at finite $x$\cite{Urbano:2010p34275}, so it is worth emphasizing that the two order parameters are determined from the same diffraction data, although their refined values are not coupled; the magnetic moment is determined by the integrated intensity of the magnetic Bragg peaks and the orthorhombicity is determined by the splittings of structural Bragg peaks. We can therefore draw two unambiguous conclusions from the data. First, the transition temperatures for both structural and antiferromagnetic order are identical and, second, that the two order parameters are strongly coupled. 

When the two transitions are coincident, they are predicted to be first-order in Ginzburg-Landau treatments of the magnetoelastic coupling\cite{Bergman:1976p36257,Barzykin:2009p21345,Cano:2010p34098}. Cano \textit{et al} show that a linear-quadratic magnetoelastic coupling generates an effective shear stress in the magnetically ordered phase\cite{Cano:2010p34098}, driving a structural distortion if \Ts\ would fall below \TN\ in the absence of coupling. When the uncoupled \Ts\ is greater than \TN, as in \BFCA\ and other transition-metal-doped compounds, the two transitions can be distinct\cite{Canfield:2010p34239}. 

On the other hand, the fact that $M\propto\delta$ in \BKFA\ implies a biquadratic coupling\cite{Wilson:2009p23312}.  It is unclear why the linear-quadratic term is not relevant but, as a consequence, neither order parameter can be considered as secondary to the other. Wilson \textit{et al} proposed that the unusual coupling was due to the accidental proximity to a tetracritical point\cite{Wilson:2009p23312}, but our data show that it persists over an extended region of the phase diagram. This suggests that there may be a deeper connection between the two order parameters, as proposed, for example, by Cvetkovic and Tesanovic who postulate the existence of a "mother" instability driving a combined spin/charge/orbital-density-wave\cite{Cvetkovic:2009p24266}.

\begin{figure}[tb]
\centering
\includegraphics[width=\columnwidth]{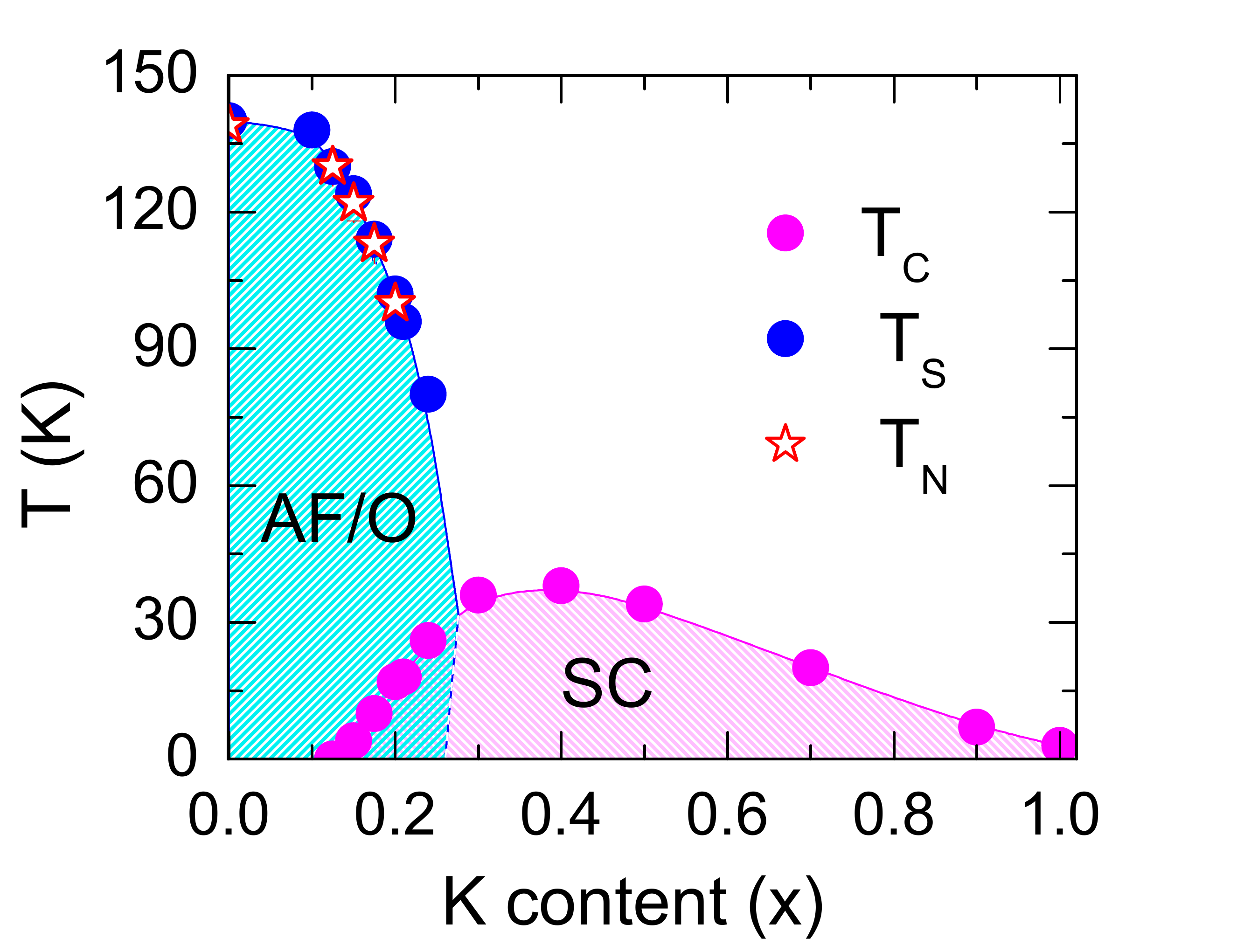}
\vspace{-0.2in}
\caption{Phase diagram of \BKFA\ with the superconducting critical temperatures (\Tc) and N\'eel temperatures (\TN), determined from magnetization measurements, and the combined antiferromagnetic/orthorhombic (AF/O) transition temperatures (\Ts), determined from neutron diffraction. Solid lines are guides to the eye. The phase boundary separating the mixed AF/O-superconducting phase from the purely superconducting phase, shown by the dotted line, is not known experimentally, but is illustrated with a positive slope as discussed in the text. 
\label{Fig5} }
\vspace{-0.2in}
\end{figure}

The complete phase diagram, compiled from both the neutron diffraction and magnetization data, is shown in Fig. 5, where we note that the error bars are all smaller than the size of the points. The antiferromagnetic/orthorhombic (AF/O) phase overlaps with superconductivity from $x=0.133$ to $\sim 0.3$. We do not currently have any measurements between $0.24\leqslant x \leqslant 0.3$ so the precise nature of the mixed phase boundary still needs to be determined. Nevertheless, we note that there is clear evidence at both $x=0.21$ (Fig. 4) and 0.24 (not shown) that there is a slight depression of the structural and magnetic order parameters on entering the superconducting phase (Fig. 4 inset). Although the statistical accuracy of the magnetic order parameter is not sufficient on its own, the orthorhombic order parameter is measured with much higher precision and shows that the biquadratically-coupled order parameters compete with the superconducting order parameter within the superconducting phase. This issue was addressed by Fernandes \textit{et al} where they point out that such competition implies that the phase boundary within the superconducting phase must have a positive slope\cite{Fernandes:2010p32347}. This has been drawn schematically in Fig. 5, although the exact slope has not been determined experimentally.

Finally, the coupling of the two order parameters throws light on the nature of the phase coexistence. The magnetization data in Fig. 1 shows that we have bulk superconductivity in all samples for $x\geqslant 0.15$, whereas the neutron diffraction data shows that the decrease in the AF/O order parameter on entering the superconducting phase is less than 5\%. This is clearly inconsistent with a mesoscopic phase separation, which would imply a significant reduction in the volume fraction of the AF/O phase below \Tc. Our results are much more consistent with the microscopic phase coexistence inferred in \BFCA. The discrepancy with earlier NMR and $\mu$SR data could be a result of improved control over chemical homogeneity in the current samples, although we will have to repeat the local probe measurements on our own samples to confirm this.

In summary, we have determined the phase diagram of \BKFA\ using high resolution neutron powder diffraction and SQUID magnetization measurements.  The magnetic and structural phase transitions at low doping are coincident and first-order, with a strong biquadratic coupling of the magnetic structure to the nuclear lattice. This unusual form of magnetoelastic coupling across an extended region of the phase diagram, including within the superconducting phase, may indicate that both order parameters are more strongly coupled than implied by conventional theories of spin density waves and orbital order\cite{Cvetkovic:2009p24266}.

We acknowledge valuable discussions with I. Paul and A. Cano. Work supported by U.S. Department of Energy, Office of Science, Office of Basic Energy Sciences, under contract No. DE-AC02-06CH11357.

\bibliography{BaKFe2As2phasediagram}

\end{document}